\documentclass[letterpaper,conference,10pt]{IEEEtran}
\usepackage[letterpaper, left=0.75in, right=0.75in, bottom=1in, top=0.75in]{geometry}
\usepackage[level=3]{latexInclusion/message}  
\usepackage{amsmath,amssymb,graphicx,mathtools,stackengine,pstool,cite}
\usepackage{acronym}  
\usepackage{color}
\usepackage[usenames,dvipsnames]{xcolor}
\usepackage{colortbl}
\usepackage{dblfloatfix}
\usepackage{latexInclusion/notation}
\usepackage{enumitem} 
\usepackage{algcompatible} 
\usepackage{algorithm} 
\usepackage{cite}
\usepackage{flushend}
\usepackage[yyyymmdd,hhmmss]{datetime} 
\newdateformat{monthyeardate}{\monthname[\THEMONTH] \THEDAY, \THEYEAR} 

\newcommand{\bd}{\begin{description}}
\newcommand{\ed}{\end{description}}
\newcommand{\be}{\begin{enumerate}}
\newcommand{\ee}{\end{enumerate}}
\newcommand{\bi}{\begin{itemize}}
\newcommand{\ei}{\end{itemize}}
\newcommand{\bl}{\begin{list}}
\newcommand{\el}{\end{list}}
\newcommand{\bt}{\begin{tabbing}}
\newcommand{\et}{\end{tabbing}}

\definecolor{BLUE}{rgb}{0,0,1}

\usepackage{acronym}

\acrodef{LiDAR}[LiDAR]{Light Detection and Ranging}%
\acrodef{PDF}[PDF]{probability density function}%
\acrodef{PMF}[PMF]{probability mass function}%
\acrodef{DA}[DA]{data association}%
\acrodef{RMSE}[RMSE]{root-mean-square error}%
\acrodef{iid}[iid]{independent and identically distributed}%
\acrodef{AWGN}[AWGN]{additive white Gaussian noise}%
\acrodef{MAP}[MAP]{maximum a posteriori}%
\acrodef{ML}[ML]{maximum likelihood}%
\acrodef{2D}[2D]{two-dimensional}%
\acrodef{3D}[3D]{three-dimensional}%
\acrodef{MTT}[MTT]{multitarget tracking}%
\acrodef{MAP}[MAP]{maximum a posteriori}%
\acrodef{MMSE}[MMSE]{minimum mean square error}%
\acrodef{ADAS}[ADAS]{advanced driver assistance systems}%
\acrodef{AUV}[AUV]{autonomous underwater vehicles}%
\acrodef{SLAM}[SLAM]{simultaneous localization and mapping}%
\acrodef{JPDA}[JPDA]{joint probabilistic data association}%

\acrodef{uwb}[UWB]{ultra-wideband}
\acrodef{fy}[FY]{fiscal year}

\acrodef{iot}[IoT] {Internet of Things}
\newacro{aws}[AWS]{Amazon Web Services}

\acrodef{ble}[BLE]{Bluetooth Low Energy}

\newacro{dfe}[DFE]{decision feedback equalizer}
\newacro{lms}[LMS]{least means square}
\newacro{ann}[ANN]{artificial neural network}
\newacro{alu}[ALU]{arithmetic logic unit}
\newacro{api}[API]{application programming interface}
\newacro{asic}[ASIC]{application-specific integrated circuit}

\newacro{sdn}[SDN]{software defined network}
\newacro{hca}[HCA]{heterogeneous computing architecture}

\acrodef{gps}[GPS]{Global Positioning System}
\acrodef{imu}[IMU]{inertial measurement unit}
\acrodef{uav}[UAV]{unmanned aerial vehicles}

\acrodef{csi}[CSI]{channel state information}
\acrodef{snr}[SNR]{signal-noise-ratio}

\acrodef{los}[LOS]{line-of-sight}
\acrodef{nlos}[NLOS]{non-line-of-sight}
\newacro{toa}[TOA]{time-of-arrival}
\newacro{tdoa}[TDOA]{time-difference-of-arrival}
\newacro{aoa}[AOA]{angle-of-arrival}
\newacro{rss}[RSS]{received signal strength}

\newacro{crb}[CRB]{Cram\'{e}r-Rao Bound}
\newacro{speb}[SPEB]{squared position error bound}
\newacro{fim}[FIM]{Fisher information matrix}
\newacro{pocs}[POCS]{projection onto convex sets}

\newacro{eed}[EED]{Engineering Every Day}
\newacro{stem}[STEM]{science, technology, engineering, and mathematics}

\DeclareMathOperator*{\argmax}{arg\,max}

\newcommand{\ist}{\hspace*{.3mm}}
\newcommand{\rmv}{\hspace*{-.3mm}}
\newcommand{\nn}{\nonumber}
\newcommand{\T}{\mathrm{T}}

\newcommand{\paperTitle}{\vspace{0.25in}Probabilistic Scan Matching:\\Bayesian Pose Estimation from Point Clouds}

\begin{document}

{
\twocolumn

\title{\paperTitle\vspace{-1mm}
}


\vspace{-1mm}
\author{\IEEEauthorblockN{Rico Mendrzik$^*$ and Florian Meyer$^{\dagger}$\hspace{-0.8mm}	
}
\\$^*$Ibeo Autmotive Systems GmbH, Hamburg, Germany
\\$^{\dagger}$University of California San Diego, San Diego, USA
\vspace{-.5mm}
\\[0.1em]
Email: rico.mendrzik@ibeo-as.com, flmeyer@ucsd.edu \vspace{-3mm}}

\maketitle

\setcounter{page}{1}

\begin{abstract}
Estimating position and orientation change of a mobile platform from two consecutive point clouds provided by a high-resolution sensor is a key problem in autonomous navigation. In particular, scan matching algorithms aim to find the translation and rotation of the platform such that the two point clouds coincide. The association of measurements in point cloud one with measurements in point cloud two is a problem inherent to scan matching. Existing methods perform non-probabilistic data association, i.e., they assume a single association hypothesis. This leads to overconfident pose estimates and reduced estimation accuracy in ambiguous environments. Our probabilistic scan matching approach addresses this issue by considering all association hypotheses with their respective likelihoods. We formulate a holistic Bayesian estimation problem for both data association and pose estimation and present the corresponding joint factor graph. Near-optimum maximum a posteriori (MAP) estimates of the sensor pose are computed by performing iterative message passing on the factor graph. Our numerical study shows performance improvements compared to non-probabilistic scan matching methods that are based on the normal distributions transform (NDT)  and  implicit moving least squares (IMLS).
\end{abstract}

\acresetall		

\section{Introduction}\label{sec:intro}
High-resolution sensors for situational awareness such as \ac{LiDAR} and imaging sonar are paramount for autonomous navigation due to their unique sensing capabilities. \ac{LiDAR} sensors can provide extremely high resolution accompanied by precise depth information and are thus ideal for automated driving and \ac{ADAS} \cite{Ziebinski:C16}. Imaging sonar generates accurate echo intensity profiles of the scanned area and can enable \ac{AUV} to navigate in an unknown and possibly unstructured environment \cite{MalliosRidaoRibasMaurelliPetillot:C10}. Beyond object tracking and classification or environmental perception, high-resolution sensors can be cornerstones for ego-state estimation which can help navigating in challenging environments \cite{BiberStrasser:C03,ZhangSingh:J16,SrivatsanXuZevallosChoset:J18,Deschaud:C18,ShanEnglot:C18,TangChenNiuWangChenLiuShiHyyppae:J15,Olson:C09,MalliosRidaoRibasHernandez:J13}.  By matching two scans obtained at different time instances, information on the change of position and orientation (also known as \textit{relative pose}) of the sensor can be extracted. A simplistic example from the \ac{LiDAR} domain is depicted in Fig. \ref{fig:scenario}.

State-of-the-art scan matching approaches \cite{BiberStrasser:C03,ZhangSingh:J16,SrivatsanXuZevallosChoset:J18,Deschaud:C18,ShanEnglot:C18,TangChenNiuWangChenLiuShiHyyppae:J15,Olson:C09,MalliosRidaoRibasHernandez:J13} typically use heuristics that require a good initial guess on the relative pose of the sensor. Moreover, they rely on a key assumption: The association of scan data can be described by a single deterministic association hypothesis. This assumption can lead to poor and overconfident estimates of the relative pose if the scans contain ambiguous data or non-associable artifacts. State-of-the-art algorithms typically fail if the scan data association is erroneous, the generating environment does not allow for an unambiguous estimation of the relative pose, or scan points originate from detection errors belong to dynamic objects. Another frequently-occurring challenge of the scan matching paradigm, is partial observability of the environment, which describes the phenomenon that two scans recorded at different poses almost never depict the identical scene.

\begin{figure}[t]
	\hspace*{-1mm}	
	\centering
	\psfragfig![width=.999\columnwidth,keepaspectratio]{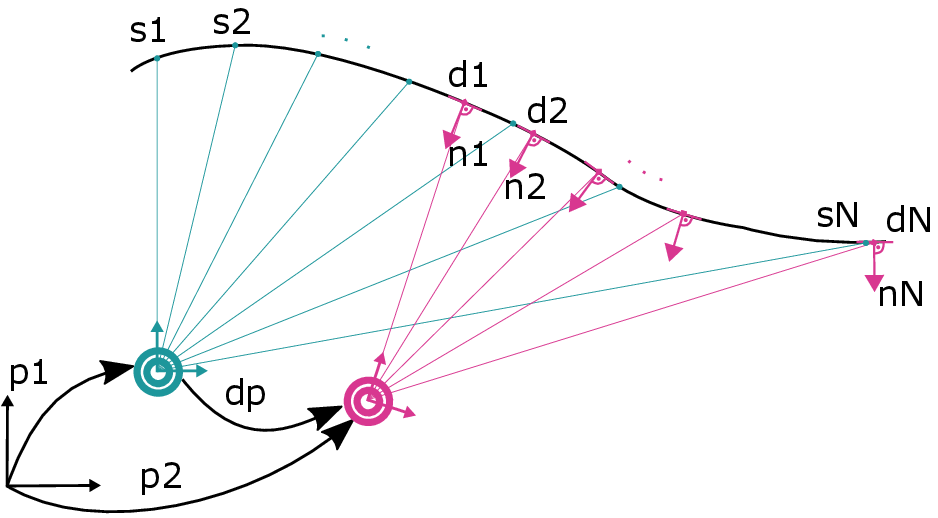}%
	\caption{Scenario: Our goal is to infer the relative pose, $\Delta\RM{P}$, of \ac{LiDAR} sensor (concentric circles) by matching the source point cloud $\V{s} = [\V{s}^{(1)\T} \ist\ist\ist \V{s}^{(1)\T} \ist\ist\ist \hdots \ist\ist\ist \V{s}^{(N_{\mathrm{S}})\T}]^\T$ (green dots) and destination surface, consisting of scan points $\V{d} = [\V{d}^{(1)\T} \ist\ist\ist \V{d}^{(2)\T} \ist\ist\ist \hdots \ist\ist\ist \V{d}^{(N_{\mathrm{D}})\T}]^\T$ (magenta diamonds) and normal vectors $\V{n} = [\V{n}^{(1)\T} \ist\ist\ist \V{n}^{(2)\T} \ist\ist\ist \hdots \ist\ist\ist \V{n}^{(N_{\mathrm{D}})\T}]^\T$ (magenta arrows). Laser rays are indicated by thin green and magenta lines originating from the \ac{LiDAR} sensor.}
	\label{fig:scenario}%
\end{figure}

In contrast to state-of-the-art method, we formulate scan matching as a Bayesian inference problem. This approach aims to increase resilience to sensing artifacts and partial observability by performing scan associations probabilistically \cite{BarShalomWillettTian:B11,WilliamsLau:J14,MeyerWin:J20,MeyWil:C20}. Our main contributions can be summarized as follows:
\begin{itemize}
	\item We derive a stochastic model which entangles both scan association and pose estimation in a holistic fashion. 
	\item We present a Bayesian inference algorithm for relative pose estimation based on probabilistic scan matching.
\end{itemize}

\textit{Notation:} Random variables are displayed in sans serif, upright fonts; their realizations in serif, italic fonts. Vectors and matrices are denoted by bold lowercase and uppercase letters, respectively. For example, a random variable and its realization are denoted by $\rv{x}$ and $x$; a random vector and its realization are denoted by $\RV{x}$ and $\V{x}$. Sets are denoted by calligraphic font, e.g. $\Set{X}$. Furthermore, ${\V{x}}^{\text T}$ denotes the transpose of vector $\V x$. $f(\V x)$ denotes the \ac{PDF} or \ac{PMF} of continuous or discrete random vector $\RV x$.
Finally, $\V{0}_n$ is the row vector of size $n$ containing all zeros.
\vspace{1mm}

\section{Fundamentals and Problem Formulation}
We define the notion of poses, describe our sensor and measurement models, and present our problem formulation.

\subsection{Preliminaries}
\vspace{-1mm}
\subsubsection{Pose Representation}
  We refer to the  position and orientation of a rigid object as pose. The pose of the sensor in $N_{\mathrm{dim}}$-dimensional space can be conveniently represented by the following matrix from the special Euclidean group $\textit{SE}(N_{\text{dim}})$\vspace{-.5mm}\cite{ThrunBurgardFox:B05}
\begin{equation*}
\RM{P} = 
\begin{pmatrix}
\RM{R} & \RV{t} \\
\V{0}_{N_{\text{dim}}} & 1
\end{pmatrix}\vspace{0mm}
\end{equation*}
where $\RM{R}$ is a $N_{\text{dim}} \rmv\times\rmv N_{\text{dim}}$ rotation matrix and $\RV{t} \in \mathbb{R}^{N_{\text{dim}}}$ is a translation vector. From source pose $\RM{P}_{\mathrm{s}}$ and destination pose $\RM{P}_{\mathrm{d}}$, we can determine the relative pose $\Delta\RM{P} \in \textit{SE}(N_{\text{dim}})$ \vspace{0mm}as 
\begin{equation*}
\Delta\RM{P} = \RM{P}_{\mathrm{d}} \RM{P}_{\mathrm{s}}^{-1} = 
\begin{pmatrix}
\Delta \RM{R} & \Delta \RV{t} \\
\V{0}_{N_{\text{dim}}} & 1
\end{pmatrix} 
\end{equation*}
where the $N_{\text{dim}} \rmv\times\rmv N_{\text{dim}}$ rotation matrix  $\Delta \RM{R}$ and the translation vector $\Delta\RV{t} \in \mathbb{R}^{N_{\text{dim}} }$ describe the change of orientation and displacement between $\RM{P}_{\mathrm{s}}$ and $\RM{P}_{\mathrm{d}}$, respectively.

\subsubsection{Sensor Characteristics}
We consider a high-resolution sensors that illuminates the environment and uses reflected and scattered signal components to infer spatial information on the observed scene. Typically, sets of scan points (so-called point clouds) are the results of sensing. Here, we consider two distinct sets: \textit{source} and \textit{destination } point clouds. We define the destination point cloud as 
\begin{equation*}
\V{d} \triangleq [ \V{d}^{(1)\T} \ist\ist\ist \V{d}^{(2)\T} \ist\ist\ist \hdots \ist\ist\ist \V{d}^{(N_{\mathrm{D}})\T} ]^\T\rmv.
\end{equation*}
where $\V{d}^{(l)} \in \mathbb{R}^{N_{\text{dim}}}$ denotes the $l^{\text{th}}$ noisy scan point in Cartesian coordinates. Similarly,  we introduce the source point cloud $\V{s} \triangleq [ \V{s}^{(1)\T} \ist\ist\ist \V{s}^{(2)\T} \ist\ist\ist \hdots \ist\ist\ist \V{s}^{(N_{\mathrm{S}})\T} ]^\T.$ 

\subsubsection{Surface Information}
Surface normal vectors are a common method to preserve information on the shapes of the underlying surfaces without the need of having explicit representations of them \cite{CarrBeatsonCherrieMitchellFrightMcCallumEvans:C01,DemantkeMalletDavidVallet:J12,KlasingAlthoffWollherrBuss:C09}. We build upon this idea and augment every destination scan point $\V{d}^{(i)}$ by a normal vector $\V{n}^{(i)} \in \mathbb{R}^{N_{\mathrm{dim}}},i= 1,2,...,N_{\mathrm{D}}$. The normal vector $\V{n}^{(i)}$ of destination scan point $\V{d}^{(i)}$ is determined based on the set of neighboring destination scan points $\Set{N}_{\mathrm{D}}^{(i)} = \{\V{d}^{(i')} |  \| \V{d}^{(i')} - \V{d}^{(i)}\| \leq d_{\mathrm{TH}}  \}$ where the threshold $d_{\mathrm{TH}}$ controls the size of the neighborhood. The literature on the computation of surface normal vectors is rich and the interested reader can find an overview in \cite{KlasingAlthoffWollherrBuss:C09}. In our work, we determine $i^{\mathrm{th}}$ surface normal vector, $\V{n}^{(i)}$, as the eigenvector which corresponds to the smallest eigenvalue of the sample covariance matrix of $\Set{N}^{(i)} $. Hereafter, we will refer to destination scan points and their associated normal vector as destination surface points. 

\subsection{Measurement Model}
Before introducing our measurement model, it is worth noting that all scan points are considered in homogeneous coordinates to cope with our pose representation, i.e., $\V{d}^{(i)}_{\mathrm{H}} \rmv\rmv \triangleq \rmv\rmv [\V{d}^{(i)\T} \ist\ist\ist 1]^{\T}$, $\V{n}^{(i)}_{\mathrm{H}}  \rmv\rmv \triangleq \rmv\rmv [\V{n}^{(i)\T} \ist\ist\ist 0]^{\T},i \rmv\rmv = \rmv\rmv 1,2,\hdots, N_{\mathrm{D}}$,  and $\V{s}^{(j)}_{\mathrm{H}}  \rmv\rmv \triangleq \rmv\rmv [\V{s}^{(j)\T} \ist\ist\ist 1]^{\T}, j \rmv\rmv = \rmv\rmv  1,2,\hdots, N_{\mathrm{S}}$. For mathematical convenience, we introduce stacked vectors of source points, destination points and their normals as follows: $\V{s}_{\mathrm{H}} = [\V{s}_{\mathrm{H}}^{(1)\T} \ist\ist\ist \V{s}_{\mathrm{H}}^{(2)\T} \ist\ist\ist \hdots \ist\ist\ist \V{s}_{\mathrm{H}}^{(N_{\mathrm{S}})\T}]^\T$, 
$\V{d}_{\mathrm{H}} = [\V{d}_{\mathrm{H}}^{(1)\T} \ist\ist\ist \V{d}_{\mathrm{H}}^{(2)\T} \hdots \ist\ist\ist \V{d}_{\mathrm{H}}^{(N_{\mathrm{D}})\T}]^\T$\rmv\rmv,  and $\V{n}_{\mathrm{H}} = [\V{n}_{\mathrm{H}}^{(1)\T} \ist\ist\ist \V{n}_{\mathrm{H}}^{(2)\T} \ist\ist\ist \hdots \ist\ist\ist \V{n}_{\mathrm{H}}^{(N_{\mathrm{D}})\T}]^\T$.

\subsubsection{Point-to-plane Error Metric}
We can link the relative pose $\Delta \RM{P}$ to source scan points, $\V{s}^{(j)},\forall j$, and destination surface points, $\{\V{d}^{(i)},\V{n}^{(i)}\},\forall i$, via a point-to-plane metric \cite{Low:TR04}. In particular, our measurement model is given by
\begin{equation}
\V{n}^{(i)}_{\mathrm{H}} \left( \V{d}^{(i)}_{\mathrm{H}} - \Delta \RM{P} \V{s}^{(j)}_{\mathrm{H}} \right) = e^{(j)} \quad j = 1,2,\hdots, N_{\mathrm{S}}
\label{eq:point2plane}
\end{equation}
in which $e^{(j)}$ is an error term with known\footnote{In partice, full knowledge of $f\big(e^{(j)}\big)$ is typically unavailable and the \ac{PDF} needs to be approximated by its most significant moments.} \ac{PDF} $f\big(e^{(j)}\big)$. The left-hand side of \eqref{eq:point2plane} is typically referred to as pseudo measurements \cite{Richards:C95}. For future reference, we also introduce the joint measurement \vspace{.3mm} vectors $\V{z}^{(i,j)} \rmv\triangleq\rmv [ \V{d}_{\mathrm{H}}^{(i)\T}\ist\ist\ist \V{n}_{\mathrm{H}}^{(i)\T} \ist\ist\ist \V{s}_{\mathrm{H}}^{(j)\T} ]^\T\rmv\rmv,$ $\V{z}^{(i)} = [\V{d}_{\mathrm{H}}^{(i)\T}$ $\V{n}_{\mathrm{H}}^{(i)\T} \ist\ist\ist\ist\ist \V{s}_{\mathrm{H}}^{\T}]^{\T}\rmv\rmv$, and $\V{z} = [\V{d}_{\mathrm{H}}^{\T} \ist\ist\ist \V{n}_{\mathrm{H}}^{\T} \ist\ist\ist\ist\ist \V{s}_{\mathrm{H}}^{\T}]^{\T}\rmv$. Note \vspace{.2mm} that from \eqref{eq:point2plane} and the error statistics $f(e^{(j)})$, we can directly obtain conditional \acp{PDF} $f(\V{z}^{(i,j)}|\Delta\M{P})$ for each pair $(i,j).$

 The measurement model \eqref{eq:point2plane} is briefly discussed in what follows. Suppose we knew the relative pose $\Delta \RM{P}$ and the association $(i,j)$ of source point $\V{s}^{(j)}_{\mathrm{H}}$ and destination surface point $\{\V{d}^{(i)}_{\mathrm{H}},\V{n}^{(i)}_{\mathrm{H}}\}$. Then, the expression within the parenthesis on the left-hand side of  \eqref{eq:point2plane} implies that we can transform the source scan point $\V{s}^{(j)}_{\mathrm{H}}$ via relative pose $\Delta \RM{P}$ and we will retrieve the destination scan point $\V{d}^{(i)}_{\mathrm{H}}$. Note that the sensor does not measure the exact same point on a surface from different poses due to its finite resolution (see Fig. \ref{fig:scenario}). We account for this by embedding the local information of the surface. This is achieved by multiplying with the surface normal $\V{n}^{(i)}_{\mathrm{H}}$. Thus, the deviation of the transformed source point from destination point is measured only in the direction of the surface normal. The remaining error is given by  $e^{(j)}$.


\subsubsection{Associability of Scan Points}
High-resolution sensing, with effects such as occlusion and finite sensing range, make scenes appear differently even if they are observed from similar poses. As a result, the number of source points $N_{\mathrm{S}}$ and destination points $N_{\mathrm{D}}$ are generally different and not all scan points are associable. For robust scan matching, we need to account for scan points which cannot be associated. In particular, we introduce the destination-oriented data association vector $\RV{a} = [\rv{a}^{(1)} \ist\ist\ist \rv{a}^{(2)} \ist \hdots \ist \rv{a}^{(N_{\mathrm{D}})}]^{\T}$ in which $\rv{a}^{(i)},i \in \{ 1,2,\hdots, N_{\mathrm{D}}\}$ refers to the association variable of the $i^{\mathrm{th}}$ destination surface point. $\rv{a}^{(i)}$ can take any value from the set $\{ 0,1,\hdots, N_{\mathrm{S}} \}$, where $\rv{a}^{(i)} = j$ means that the $i^{\mathrm{th}}$ destination surface point relates to the $j^{\mathrm{th}}$ source point. Moreover, $\rv{a}^{(i)} = 0$ means that the destination surface point $\{\V{d}_{\mathrm {H}}^{(i)},\V{n}_{\mathrm {H}}^{(i)}\}$ is not associable to any source point $\V{s}_{\mathrm{H}}^{(j)},\forall j$. 
\vspace{1mm}



\subsection{Problem Formulation}
Fig. \ref{fig:scenario} graphically depicts the problem we attempt to solve. Our goal is to infer the relative pose $\Delta\RM{P}$ of the sensor based on the source and destination point clouds by transforming the source point cloud onto the destination point cloud. Note that the associations between source scan points and destination surface points $\RV{a}$ are unknown and need to be inferred jointly with the relative pose  $\Delta\RM{P}$ which is a random matrix on the manifold $\textit{SE}(N_{\text{dim}})$.

	
In the pursuit of solving this problem robustly, we resort to the probabilistic domain. Our goal is to obtain a Bayes-optimum estimate of the relative pose $\Delta\RM{P}$ considering all valid data association hypotheses $\RV{a}$. First, we marginalize the joint posterior distribution $f(\Delta \M{P},\V{a}|\V{z})$ to obtain $f(\Delta \M{P}|\V{z})$ which factors in all valid data association hypotheses. Then, we derive an \ac{MAP} estimate by maximizing the marginal posterior distribution, i.e.,
\begin{equation}
\Delta\hat{\M{P}}_{\mathrm{MAP}} = \argmax_{\Delta\M{P}} f(\Delta\M{P}|\V{z}).
\vspace{-1.2mm}
\label{eq:estimationAgent}
\end{equation}
Note that determining $f(\Delta \M{P}|\V{z})$ by direct marginalization $\big(\text{i.e., }f(\Delta \M{P}|\V{z})= \sum_{\V{a}}f(\Delta \M{P},\V{a}|\V{z})\big)$ is infeasible due to the high-dimensionality of the summation space as discussed in Section~\ref{sec:stoMod}. We consider feasible approximate calculation based on the notion of message passing \cite{LoeligerDauwelsHuKorlPingKschischang:J07,Loeliger:J04,KschischangFreyLoeliger:J01}.

\vspace{0mm}
\section{Stochastic Modeling}
\label{sec:stoMod}
Here, we present our novel stochastic model which couples data association, sensor pose, and point cloud measurements.


\vspace{-1mm}
\subsection{Modeling of the Point-to-Surface-Point Associations}
\vspace{-1mm}
We attempt to associate $N_{\mathrm{S}}$ source points to $N_{\mathrm{D}}$ destination surface points. Recall, though, that not all points in the source and destination scan are associable.   We assume the number of non-associable source points, $\rv{N}_{\mathrm{NA}}$, to be random and Poisson distributed with mean $\lambda_{\mathrm{NA}}$. Hence the total number of source points, $\rv{N}_{\mathrm{S}}$, is also random and given by
\begin{equation}
\rv{N}_{\mathrm{S}} = \rv{N}_{\mathrm{NA}} +  | \Set{D}_{\RV{a}} |,
\label{eq:numSource}
\end{equation}
where $\Set{D}_{\RV{a}} \triangleq \{ i\in \{1,2,\hdots, N_{\mathrm{D}}\}: \rv{a}^{(i)}\neq 0 \}$.

Valid associations are those in which each source point $\V{s}_{\mathrm{H}}^{(j)}$ is associated to at most one destination surface point $\{\V{d}_{\mathrm{H}}^{(i)},\V{n}_{\mathrm{H}}^{(i)}\}$, and
 each destination surface point $\{\V{d}_{\mathrm{H}}^{(i)},\V{n}_{\mathrm{H}}^{(i)}\}$ is associated to at most one source point $\V{s}_{\mathrm{H}}^{(j)}$. This assumptions imposes constraints on $\RV{a}$. In particular, $\RV{a}$ is valid if and only if $\rv{a}^{(i)}=j \in \{1,\hdots, N_{\mathrm{S}}  \}$, $ \rv{a}^{(i')}\neq j, \forall i \neq i'$. These constraints can be conveniently expressed by
\begin{equation}
\Psi(\V{a}) = 
\begin{cases}
0,\quad & a^{(i)} = a^{(i')}=j, i\neq i', j\neq 0
\\
1,\quad & \text{otherwise}.
\end{cases}
\end{equation} 
Note that the cardinality of valid hypothesis is huge, rendering enumeration of all hypotheses intractable  \cite{WilliamsLau:J14,MeyerKropfreiterWilliamsLauHlawatschBracaWin:J18}. To address this issue, we introduce a redundant formulation of $\RV{a}$: the source-oriented data association vector $\RV{b} = [\rv{b}^{(1)} \ist\ist\ist \rv{b}^{(2)} \ist\ist\ist \hdots \ist\ist\ist \rv{b}^{(N_{\mathrm{S}})}]^{\T}$, in which $\rv{b}^{(j)} \in \{1,\hdots,N_{\mathrm{D}}\}$ and $\rv{b}^{(j)} = i $ means that the $j^{\mathrm{th}}$ source point is associated to the $i^{\mathrm{th}}$ destination surface point. If $\rv{b}^{(j)}=0$, the $j^{\mathrm{th}}$ source point is non-associable. For source-oriented data association vectors $\RV{b}$ we have $\rv{b}^{(j)}=i \in \{1,\hdots, N_{\mathrm{D}} \}$, $ \rv{b}^{(j')}\neq i, \forall j' \neq j$. Note that the set of all valid vectors $\RV{b}$ contains exactly the same information as that of all valid vectors $\RV{a}$. 

The reason for introducing this redundant data association representation is that validity of association hypotheses can now be described via pairs of $\rv{a}^{(i)}$ and $\rv{b}^{(j)}$ instead of the entire vector $\RV{a}$, making evaluation tractable. This is described in more detail in \cite{WilliamsLau:J14,MeyerKropfreiterWilliamsLauHlawatschBracaWin:J18}. It can be easily verified that valid data associations can be described by 
\begin{equation}
\Psi(\V{a},\V{b}) = \prod_{i=1}^{N_{\mathrm{D}}}\prod_{j=1}^{N_{\mathrm{S}}}\psi_{i,j}(a^{(i)},b^{(j)}),
\label{eq:psiAB}
\vspace{-2mm}
\end{equation}
where
\begin{equation}
\vspace{-1mm}
\psi_{i,j}(a^{(i)},b^{(j)}) \triangleq
\begin{cases}
0,\quad & a^{(i)}=j, b^{(j)}\neq i 
\\
& b^{(j)} = i, a^{(i)}\neq j
\\
1,\quad & \text{otherwise}.
\end{cases}
\label{eq:indicator}
\end{equation}
Note that we treat both $\RV{a}$ and $\RV{b}$ as random vectors that are inferred jointly with the relative pose $\Delta \RM{P} $.
\begin{figure*}[!t]
	\normalsize
	\setcounter{equation}{8}
	\begin{equation}
		f(\Delta \M{P},\V{a},\V{b}|\V{z})  \propto  f(\Delta\M{P}) v(\Delta\M{P};\V{s}_{\mathrm{H}}) \prod_{i=1}^{N_{\mathrm{D}}} q_i(\Delta\M{P},a^{(i)};\V{z}^{(i)})  \prod_{j=1}^{N_{\mathrm{S}}} \psi_{i,j}(a^{(i)},b^{(j)}) 
		\label{eq:jointPosterior}
	\end{equation}
	\setcounter{equation}{6}
	\hrulefill
	\vspace{-4mm}
\end{figure*}

\begin{figure}[!t]
	\vspace{-2mm}
	\centering
	\psfragfig![width=1\columnwidth,keepaspectratio]{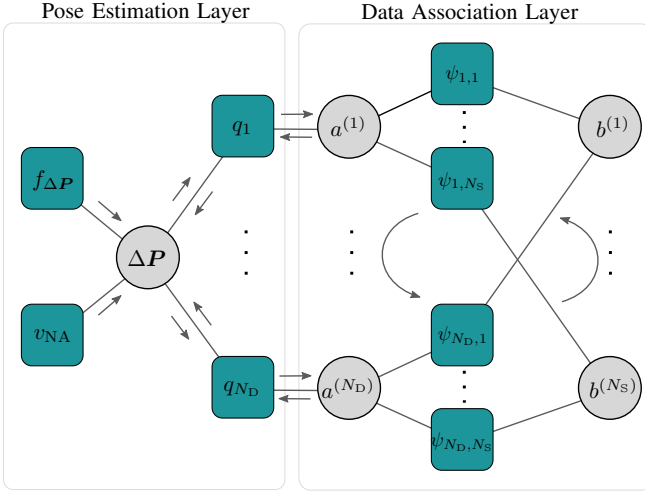}%
	\vspace*{-2mm}
	\caption{Factor graph of the joint posterior distribution in \eqref{eq:jointPosterior}. The graph can be sub-divided into a data association layer and a pose inference layer. In the latter, our goal is to infer the marginal posterior distribution $f(\Delta\M{P}|z)$ via message passing (messages are indicated with grey arrows). This \ac{PDF} will maximized in a subsequent step to obtain an \ac{MAP} estimate of the relative pose $\Delta\RM{P}$. We use the following shorthand notations within the graph: $f_{\Delta\M{P}} \triangleq f(\Delta\M{P})$, $q_i \triangleq q_i(\Delta\M{P},a^{(i)};\V{z}^{(i)})$, $\psi_{i,j} \triangleq \psi_{i,j}(a^{(i)},b^{(j)})$, $v_{\mathrm{NA}}\triangleq v(\Delta\M{P};\V{s}_{\mathrm{H}})$.}
	\label{fig:fator_graph}%
	\vspace{2mm}
\end{figure}

\vspace{-1mm}
\subsection{Joint Posterior Distribution}
In what follows, we derive the joint posterior distribution $f(\Delta \M{P},\V{a},\V{b}|\V{z})$ and the corresponding factor graph. For $\V{z}$ observed and thus fixed, we obtain
\begin{equation}
f(\Delta \M{P},\V{a},\V{b}|\V{z}) \propto f(\V{z},\V{a},\V{b},N_{\mathrm{S}}|\Delta \M{P}) f(\Delta \M{P}).
\label{eq:jointPosterior1}
\vspace{0mm}
\end{equation}
$f(\Delta \M{P})$ is a random matrix pdf with support $\textit{SE}(N_{\text{dim}})$ that incorporates prior knowledge on the relative pose $\Delta \RM{P}$. The conditional \ac{PDF} $f(\V{z},\V{a},\V{b},N_{\mathrm{S}}|\Delta \M{P})$ is given by
\begin{align}
\vspace{-4mm}
\label{eq:conditionalPDF}
&f(\V{z},\V{a},\V{b},N_{\mathrm{S}}|\Delta \M{P}) \\[1mm] \nn
&=  \!\frac{e^{-\lambda_{\mathrm{NA}}}}{N_{\mathrm{NA}}!} \Psi(\V{a},\V{b}) v(\Delta\M{P};\V{s}_{\mathrm{H}}) \hspace{-.4mm} \prod_{i = 1}^{N_{\mathrm{D}}} q_i(\Delta\M{P},a^{(i)};\V{z}^{(i)})
\end{align}
where we introduced $v(\Delta\M{P};\V{s}_{\mathrm{H}}) \triangleq \prod_{j=1}^{N_{\mathrm{S}}} f_{\mathrm{NA}}(\V{s}_{\mathrm{H}}^{(j)} | \Delta\M{P})$ and
\begin{align}
\setcounter{equation}{9}
\hspace{-5mm} q_i(\Delta\M{P},a^{(i)};\V{z}^{(i)}) \nn& 
\\
&
\hspace{-15mm}= 
\begin{cases}
f_{\mathrm{A}}(\Delta \M{P}) \frac{f(\V{z}^{(i,a^{(i)})}|\Delta\M{P}) }{\lambda_{\mathrm{NA}} f_{\mathrm{NA}}(\V{s}_{\mathrm{H}}^{(a^{(i)})}|\Delta \M{P})} \quad &  a^{(i)} \neq 0
\\
1 - f_{\mathrm{A}}(\Delta \M{P}) \quad & a^{(i)} = 0
\end{cases} 
\end{align}  
and used $\Psi(\V{a},\V{b})$ from \eqref{eq:psiAB}. A detailed derivation of \eqref{eq:conditionalPDF} together with the underlying assumptions is provided in Appendix \ref{app:derivationConditionalPDF}. By using \eqref{eq:conditionalPDF} in \eqref{eq:jointPosterior1}, we obtain the joint posterior distribution $f(\Delta\M{P},\V{a},\V{b}|\V{z})$ shown in \eqref{eq:jointPosterior}.

Let us discuss the expression in \eqref{eq:jointPosterior} to gain further insights into our model. Suppose that we have a destination surface point $i$, $\{\RV{d}_{\mathrm{H}}^{(i)},\RV{n}_{\mathrm{H}}^{(i)}\}$, that is associated with source point $\RV{s}_{\mathrm{H}}^{(j)}$, i.e., $a^{(i)} = j$. Then, we have $q_i(\Delta\M{P},j,\V{z}^{(i)}) = f_{\mathrm{A}}(\Delta\M{P})f(\V{z}^{(i,j)}|\Delta\M{P}) / (\lambda_{\mathrm{NA}} f_{\mathrm{NA}}(\V{s}_{\mathrm{H}}^{(j)}|\Delta \M{P}))$, where $f_{\mathrm{A}}(\cdot)$ accounts for the fact that destination surface point $i$ is associable and contributes pseudo measurement $\V{z}^{(i,j)}$ with likelihood $f(\V{z}^{(i,j)}|\Delta\M{P})$. Lastly, the term in the denominator, $ f_{\mathrm{NA}}(\V{s}_{\mathrm{H}}^{(j)}|\Delta \M{P})$, reflects the fact that source point $j$, $\V{s}_{\mathrm{H}}^{(j)}$, is associable as well since it cancels the $j^{\text{th}}$ term in $v(\Delta\M{P};\V{s}_{\mathrm{H}})$. On the contrary, if the $i^{\mathrm{th}}$ destination surface point is not associated with any source point, we have $q_i(\Delta\M{P},0,\V{z}^{(i)}) = 1 - f_{\mathrm{A}}(\Delta\M{P})$. This term takes the likelihood of non-associability, $1 - f_{\mathrm{A}}(\Delta\M{P})$, of destination surface point $i$ into account. Finally, $\Psi(\V{a},\V{b})$ enforces validity of associations.
\begin{algorithm}[!t]
	\caption{Obtaining the Marginal Posterior $f(\Delta\M{P}|\V{z})$}\label{alg:messagePassing}	
	\begin{algorithmic}[1]
		\STATE Initialize prior information 
		\STATEx \vspace{-6mm} \begin{align*}  \mu_{f_{\Delta\M{P}} \rightarrow \Delta\M{P} }(\Delta\M{P}) &= f(\Delta\M{P}) 
			\\ 
			\mu_{v_{\mathrm{NA}} \rightarrow \Delta\M{P} }(\Delta\M{P}) &= v(\Delta\M{P}; \V{s}_{\mathrm{H}}) \end{align*} \vspace{-6mm}
		
		\STATE Forward prior information
		\STATEx \textbf{For} $i=1,2,\hdots,N_{\mathrm{D}}$
		\STATEx \vspace{-5mm} \begin{align*}  \mu_{\Delta\M{P} \rightarrow q_{i}}(\Delta\M{P}) = \mu_{f_{\Delta\M{P}} \rightarrow \Delta\M{P} }(\Delta\M{P}) \mu_{v_{\mathrm{NA}} \rightarrow \Delta\M{P} }(\Delta\M{P}) \end{align*}\vspace{-5mm}
		\STATEx \textbf{End For}
		
		\STATE Evaluate associations 
		\STATEx \textbf{For} $i=1,2,\hdots,N_{\mathrm{D}}$
		\STATEx \vspace{-5mm} \begin{align*} \mu_{q_{i} \rightarrow a^{(i)}}(a^{(i)}) &\!=\! \rmv\rmv \int \rmv\rmv\rmv q_i(\Delta\M{P},a^{(i)};\V{z}^{(i)}) \mu_{\Delta\M{P} \rightarrow q_{i}}(\Delta\M{P})\mathrm{d}\Delta\M{P} 
			\\
			& \approx \sum^{N_{\mathrm{P}}}_{p = 1} q_i(\Delta\M{P}^{(p)},a^{(i)};\V{z}^{(i)})v(\Delta\M{P}; \V{s}_{\mathrm{H}}) \end{align*} 
		\vspace{-3mm}
		\STATEx where $\Delta\M{P}^{(p)}$, $p = 1,2,\dots,N_{\mathrm{P}}$ are random samples drawn from $\mu_{f_{\Delta\M{P}} \rightarrow \Delta\M{P} }(\Delta\M{P}) =  f(\Delta\M{P})$.
		\STATEx \textbf{End For}
		
		\STATE Perform $N_{\mathrm{DA}}$ data association iterations
		
		\STATEx \textbf{For} $l=1,2,\hdots,N_{\mathrm{DA}}$
		\STATEx \vspace{-6mm} \begin{align*} &\text{ Message exchange  between } a^{(i)} \text{ and } b^{(j)}  \text{ see \cite{WilliamsLau:J14,MeyerKropfreiterWilliamsLauHlawatschBracaWin:J18}} 
			\\ 
			& (\mu^{(l)}_{a_{i} \rightarrow \psi_{i,j}}, \mu^{(l)}_{ \psi_{i,j} \rightarrow b^{(j)}}, \mu^{(l)}_{ b^{(j)} \rightarrow \psi_{i,j}},\mu^{(l)}_{ \psi_{i,j} \rightarrow a^{(i)}})\end{align*}
		\STATEx \textbf{End For}
		
		\STATE Forward data association result
		\STATEx \textbf{For} $i=1,2,\hdots,N_{\mathrm{D}}$
		\STATEx \vspace{-5mm} \begin{align*} \mu_{a^{(i)} \rightarrow q_{i}}(a^{(i)}) = \prod_{j=1}^{N_{\mathrm{S}}}  \mu^{(N_{\mathrm{DA}})}_{\psi_{i,j} \rightarrow a_{i}}(a^{(i)})\end{align*}\vspace{-5mm}
		\STATEx \textbf{End For}
		
		\STATE Marginalize associations 
		\STATEx \textbf{For} $i=1,2,\hdots,N_{\mathrm{D}}$
		\STATEx \vspace{-5mm} \begin{align*} \mu_{q_{i} \rightarrow \Delta\M{P}}(\Delta\M{P}) = \sum_{a^{(i)} = 0}^{N_{\mathrm{S}}} q_i(\Delta\M{P},a^{(i)};\V{z}^{(i)}) \mu_{a^{(i)} \rightarrow q_{i}}(a^{(i)}) \end{align*}\vspace{-3mm}
		\STATEx \textbf{End For}
		
		\STATE Determine marginalized posterior $f(\Delta\M{P}|\V{z})$
		\STATEx \vspace{-5mm} \begin{align} \hspace{-2mm} f(\Delta\M{P}|\V{z}) \approx f(\Delta\M{P}) v(\Delta\M{P};\V{s}_{\mathrm{H}}) \prod_{i=1}^{N_{\mathrm{D}}}\mu_{q_{i} \rightarrow \Delta\M{P}}(\Delta\M{P}) \label{eq:marginalPost} \end{align}
		\vspace{-2mm}
	\end{algorithmic}
\end{algorithm}

\vspace{-1.5mm}
\section{Graph-based Inference}
\vspace{-2mm}
The factor graph of the joint posterior distribution in \eqref{eq:jointPosterior} is depicted in Fig. \ref{fig:fator_graph}. Our goal is to infer the marginal posterior distribution $f(\Delta\M{P}|\V{z}) = \sum_{\V{a}}\sum_{\V{b}}f(\Delta\M{P},\V{a},\V{b}|\V{z})$ needed for the calculation of an \ac{MAP} estimate $\Delta\hat{\M{P}}_{\mathrm{MAP}}$. We employ the notation of message passing according to the sum-product rules \cite{LoeligerDauwelsHuKorlPingKschischang:J07,Loeliger:J04,KschischangFreyLoeliger:J01} to calculate an approximation of the marginal posterior distribution $f(\Delta\M{P}|\V{z})$. The messages passed along the edges of the factor graph are indicated via grey arrows in the factor graph in Fig. \ref{fig:fator_graph}. Messages that involve integration $\int\mathrm{d}\Delta\M{P}$  over the sensor pose $\Delta\M{P}$ are approximated by sampling-based integration \cite{DoucetFreitasGordon:B01}. We present the steps of our inference method in Algorithm \ref{alg:messagePassing}.

First, we perform Algorithm \ref{alg:messagePassing} to compute an accurate approximation of the marginal posterior $f(\Delta\M{P}|\V{z})$. Then, our goal is to derive the  \ac{MAP} estimate, $\Delta\hat{\M{P}}_{\mathrm{MAP}} = \argmax_{\Delta\M{P}} f(\Delta\M{P}|\V{z})$, by maximizing \eqref{eq:marginalPost} with $N_{\mathrm{IT}}$. To that end, we employ iterative numerical solvers that exploit the structure of $\textit{SE}(N_{\text{dim}})$\cite{Boumal:B20}. As initial guess, we use the pose sample that maximizes the marginal posterior, i.e.,
\begin{equation}
\label{eq:initialGuess}
\Delta\M{P}_{\mathrm{init}} = \argmax_{\Delta\M{P}^{(p)}} f(\Delta\M{P}^{(p)}|\V{z}), \quad p \in \{ 1,2,\hdots N_{\mathrm{P}} \}
\end{equation}
where $\Delta\M{P}^{(p)}$, $p = \rmv 1,2,\dots,N_{\mathrm{P}}$ are random samples drawn from $f(\Delta\M{P})$. 

We emphasize that our approach determines the full marginal posterior distribution $f(\Delta\M{P}|\V{z})$. Obtaining an estimate by finding its mode is by no means the only information that we can extract from it. We could derive, e.g., uncertainty information on our estimate that we can use to reason about the quality of matching. Such knowledge is typically required for closing loops in \ac{SLAM} applications. Beyond scan-to-scan matching, our proposed stochastic model can be embedded into online or offline factor graph-based trajectory estimation.

The computational complexities in the pose inference and data association layers (see Fig. \ref{fig:fator_graph}) as well as for optimization are given by $\mathcal{O}(N_{\mathrm{D}}N_{\mathrm{S}}N_{*})$, $*\in\{\mathrm{P,DA,IT}\}$. To avoid high complexity, segmentation of point clouds \cite{TchapmiChoyArmeniGwakSavarese:C17} can be used to generate semantic clusters (e.g., road surface, guardrails, poles, etc.) where each cluster contains significantly fewer points than the original point cloud. Running multiple instances (one for each cluster) of our proposed algorithm in parallel reduces the complexity as $N_{\mathrm{D},i}N_{\mathrm{S},i} \ll N_{\mathrm{D}}N_{\mathrm{S}}$, where  $N_{\mathrm{D},i}$ and $N_{\mathrm{S},i}$ denote the number of destination and source points in\vspace{3mm} cluster $i$, respectively.

\begin{table}[!t]
	\vspace*{.2mm}	
	\renewcommand{\arraystretch}{1.5} 
	\begin{center}
		\resizebox{1\columnwidth}{!}{
			\begin{tabular}{  p{3cm} p{1.5cm} p{1.5cm} p{1.5cm}  }
				\hline
				\textbf{Relative Trans. Error} & \textbf{Proposed} &  \textbf{NDT} \cite{BiberStrasser:C03}  & \textbf{IMLS} \cite{Deschaud:C18} \\		
				\hline
				\rowcolor{blue!10!white} $e_{\mathrm{trans},50}$ & 1.87$\%$ &1.800$\%$  & 6.60$\%$ \\
				\rowcolor{blue!10!white}  $e_{\mathrm{trans},95}$& 7.723$\%$ & 94.65$\%$ & 28.06$\%$\\
				\hline
				\textbf{Relative Rotation Error} & \textbf{Proposed} &  \textbf{NDT} \cite{BiberStrasser:C03}  & \textbf{IMLS} \cite{Deschaud:C18} \\		
				\hline
				\rowcolor{blue!10!white} $e_{\mathrm{rot},50}$ & 0.0349$^\circ$/m & 0.0203$^\circ$/m  & 0.2487$^\circ$/m   \\
				\rowcolor{blue!10!white}  $e_{\mathrm{rot},95}$& 0.1123$^\circ$/m	 & 0.0759 $^\circ$/m & 2.2016$^\circ$/m\\				
		\end{tabular}}
		\vspace{-2mm}
	\end{center}
	\caption{Error quantiles of simulated pose estimation methods.}
	\label{tab:results}
	\vspace{-1mm}
\end{table}
\vspace{-.5mm}

\section{Numerical Evaluation}
In this section, we evaluate the our scan matching approach based on simulated data and present the results. 

\subsection{Simulation Scenario}
\vspace{0mm}
We consider a vehicle that drives in the urban environment depicted in Fig. \ref{fig:trajectory}. The vehicle moves with a constant velocity of $10$ m/s along the indicated trajectory (red line). It is equipped with a $360^{\circ}$-\ac{LiDAR} sensor that generates scans every $\Delta T \rmv= \rmv80$ ms. We treat the current scan as the source scan and the previous scan as the destination scan to obtain pose estimates that are inline with the motion of the vehicle.

The sensor has an angular resolution of $1$ deg and a maximum sensing range of $100$ m. The scanner obtains range and bearing estimates that are corrupted by zero-mean Gaussian noise with standard deviation $\sigma_{\mathrm{range}}= 0.05$ m and  $\sigma_{\mathrm{bearing}}= 0.5$ deg, respectively. Non-associable source points are assumed to be uniformly distributed in range and bearing, being independent of $\Delta\RM{P}$. The number of non-associable source points is assumed Poisson distributed with mean $\lambda_{\mathrm{NA}} = 1$. For surface normal computation, we use $d_{\mathrm{TH}} = 2$ m. An exemplary scan is depicted in Fig. \ref{fig:trajectory}. We use the following parameters for our proposed algorithm. The probability of associability is constant $f_{\mathrm{A}}(\Delta\M{P}) = 0.8$. For each $j \in \{1,2,\hdots,N_{\mathrm{S}}\}$, the noise statistics of $f(e^{(j)})$ are assumed zero-mean Gaussian distributed with standard deviation $\sigma_{e^{(j)}} = 0.03$ m. Translation and rotation prior are considered uniform within the intervals $[-10,10]$ m and $[-90,90]$ $^\circ$. We performed $N_{\mathrm{MC}}=100$ Monte Carlo trials.

\subsection{Performance Comparison}
We compare our probabilistic scan matching approach to normal distributions transform (NDT) scan matching \cite{BiberStrasser:C03} and  implicit moving least squares (IMLS) scan matching \cite{Deschaud:C18}. In our implementation of IMLS, we set the neighborhood parameter to $h\rmv=\rmv2$ and obtained estimates by minimizing \cite[Eq.~(1)]{Deschaud:C18}. We employ the following error metrics to evaluate the translation and rotation drift: $e_{\mathrm{trans}} = \|\Delta \hat{\V{t}}_{\mathrm{MAP}}- \Delta\V{t}\| /\|\Delta\V{t}\|$ and $e_{\mathrm{rot}} = \arccos ((\mathrm{Tr} ( \Delta \hat{\M{R}}_{\mathrm{MAP}}^{-1} \Delta\M{R} )-1)/2 )/\|\Delta\V{t}\|$, where $\Delta\hat{\V{t}}_{\mathrm{MAP}}$ and $\Delta\hat{\M{R}}_{\mathrm{MAP}}$ are the estimated translation and rotation extracted from $\Delta\hat{\M{P}}_{\mathrm{MAP}}$, $\Delta\V{t}$ and $\Delta\M{R}$ are the ground truth translation and rotation, and $\mathrm{Tr}(\cdot)$ is the trace operator.

\begin{figure}[!t]
	\centering
	\vspace*{-1mm}	
	\hspace*{-2mm}
	\psfragfig![width=\columnwidth,keepaspectratio]{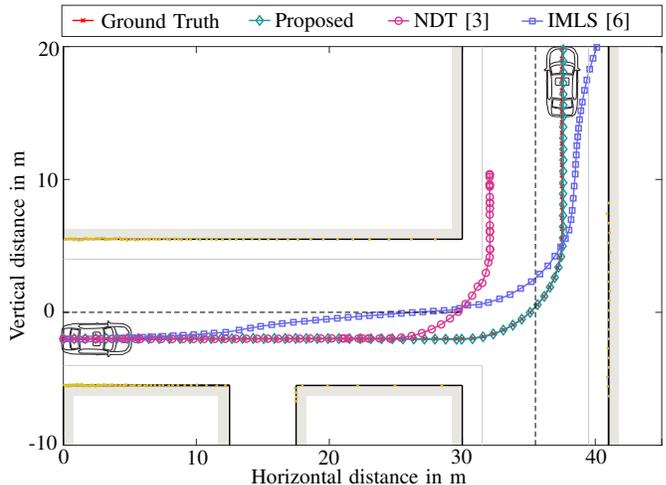}%
	\vspace*{-2mm}
	\caption{Simluation scenario and results: The ground truth trajectory is shown against the accumulated pose change estimates of the our approach and its benchmarks. Scan points obtained at the first time instance are depicted by yellow dots to visualize the characteristics of the \ac{LiDAR} sensor.} 
	\label{fig:trajectory}%
\end{figure}

Results are depicted in Tab. \ref{tab:results}. Here, $e_{\mathrm{trans}, q}$ refers to an error threshold such that $q\%$ of all relative observed translation errors $e_{\mathrm{trans}}$ are below this threshold. The same logic applies to $e_{\mathrm{rot}, q}$. For example, $e_{\mathrm{trans},50}$ is the median relative translation error. From Tab. \ref{tab:results} we observe that median relative translation and rotation errors of NDT and our proposed approach both show low drifts. However, considering $e_{\mathrm{trans}, q}$, NTD shows a significantly larger drift compared to our proposed probabilistic approach. IMLS scan matching shows large drifts, presumably due to the absence of a good initial guess provided to the optimizers. Exemplary estimated trajectories obtained via a concatenation of relative pose estimates are also shown in Fig. \ref{fig:trajectory} where we can see how closely our approach attains ground \vspace{.5mm} truth. 

\section{Conclusions}
\vspace{-.5mm}
We derived a probabilistic scan matching model which jointly describes the pose change of a high-resolution sensor along with the unknown data associations of observed scan points. Moreover, we presented an inference algorithm for that model which determines the posterior distribution of pose change given scan data. Compared to state-of-the-art scan matching algorithms, our approach considers all data association hypotheses and allows for uncertainty quantification, as the full posterior distribution is available. Numerical studies show precise pose change estimates characterized by low drift in terms of both translation and rotation. 
\vspace{-.5mm}

\appendices
\section{Derivation of $f(\V{z},\V{a},\V{b},N_{\mathrm{S}}|\Delta \M{P})$}
\label{app:derivationConditionalPDF}
\vspace{0mm}
We make the following assumptions:

\begin{enumerate}[label=A\arabic*)]
	\item Source points, $\V{s}_{\mathrm{H}}^{(j)}, \forall j$, are either associable or non-associable. Every associable source point can be associated with at most one destination surface point and every destination surface point can be associated to at most one associable source point.
	\item Valid source-to-destination associations according to A1) are uniformly distributed a priori. \label{assumption:uniformAssociations}
	\item Associability of destination points, $\RV{\theta}$, and the number of non-associable source points, $\rv{N}_{\mathrm{NA}}$, are statistically independent given the relative pose, $\Delta\RM{P}$. \label{assumption:indepdentThetaNNA}
	\item The associability of destination surface points $i$ and $i', \forall i\neq i'$, $\rv{\theta}^{(i)}$ and $\rv{\theta}^{(i')}$, are independent Bernoulli-distributed random variables given the relative pose $\Delta\RM{P}$. The probability of a destination surface point $i$ to be associable ($\rv{\theta}^{(i)}=1$) is $0\le f_{\mathrm{A}}(\Delta\M{P}) < 1, \forall i$.\label{assumption:indepdentTheta}
	\item The number of non-associable source points, $\rv{N}_{\mathrm{NA}}$, is Poisson-distributed with intensity $\lambda_{\mathrm{NA}}$. \label{assumption:poissonClutter}
	\item A non-associable source point, $\V{s}_{\mathrm{H}}^{(j)}$, is statistically independent of an associable source points, $\V{s}_{\mathrm{H}}^{(j')}$. This holds for all associable and non-associable points. \label{assumption:indepdentAssoNonAssoMeasurements}
	\item Given the relative pose $\Delta\RM{P}$, pseudo measurements, $\V{z}^{(i,j')}, \forall i,j'$, corresponding to associable source points, $\V{s}_{\mathrm{H}}^{(j')},\forall j'$, are independent of each other and distributed according to $f(\V{z}^{(i,j')}|\Delta\M{P})$. \label{assumption:indepdentAssoMeasurements}
	\item Non-associable source points, $\V{s}_{\mathrm{H}}^{(j)}, \forall j$, are \ac{iid} with \ac{PDF} $f_{\mathrm{NA}}(\V{s}_{\mathrm{H}}^{(j)}|\Delta \M{P}) \neq 0, \Delta\M{P} \in \textit{SE}(N_{\text{dim}})$. \label{assumption:clutterDistribution} 
\end{enumerate}

As a preliminary step, we derive the conditional \ac{PDF} $f(\V{z},\V{a},N_{\mathrm{S}}|\Delta \M{P})$. This is closely related to the derivation of the \ac{JPDA} filter \cite{BarShalomWillettTian:B11}. We start by decomposing $f(\V{z},\V{a},N_{\mathrm{S}}|\Delta \M{P})$ \vspace{0mm} as 
\begin{equation}
f(\V{z},\V{a},N_{\mathrm{S}}|\Delta \M{P}) = f(\V{z}|\V{a},N_{\mathrm{S}},\Delta \M{P})f(\V{a},N_{\mathrm{S}}|\Delta \M{P}).
\label{eq:conditionalPDFDer}
\end{equation}
Let us next define the vector $\RV{\theta} \triangleq [\rv{\theta}^{(1)} \ist \ist \ist \rv{\theta}^{(2)} \ist\ist\ist \hdots \ist\ist\ist \rv{\theta}^{(N_{\mathrm{D}})}]^\T$ that indicates which destination surface point $i\in \{ 1,2,\hdots,N_{\mathrm{D}} \}$ is associable ($\rv{\theta}^{(i)} = 1$) or non-associable ($\rv{\theta}^{(i)} = 0$). Note that $\rv{\theta}^{(i)} = 1$ implies $\rv{a}^{(i)} \neq 0$ and $\rv{\theta}^{(i)} = 0$ implies $\rv{a}^{(i)} = 0$. Due to \eqref{eq:numSource}, the mapping between ($\RV{a},\RV{\theta},\rv{N}_{\mathrm{NA}}$) and ($\RV{a},\rv{N}_{\mathrm{S}}$) is thus deterministic as well as bijective and the factor $f(\V{a},N_{\mathrm{S}}|\Delta \M{P})$ in \eqref{eq:conditionalPDFDer} can also be written as $f(\V{a},\V{\theta},N_{\mathrm{NA}}|\Delta \M{P}) =f(\V{a},N_{\mathrm{S}}|\Delta \M{P}) $. 

As a result of the chain rule and by using A3), we obtain
\begin{align}
&f(\V{a},N_{\mathrm{S}}|\Delta \M{P}) \nn\\
&\hspace{7mm}= f(\V{a},\V{\theta},N_{\mathrm{NA}}|\Delta \M{P}) \nn\\
&\hspace{7mm}=  f(\V{a}|\V{\theta},N_{\mathrm{NA}},\Delta \M{P}) f(\V{\theta}|\Delta \M{P},N_{\mathrm{NA}}) f(N_{\mathrm{NA}}|\Delta \M{P}) \nn\\[1mm]
&\hspace{7mm}=  f(\V{a}|\V{\theta},N_{\mathrm{NA}},\Delta \M{P}) f(\V{\theta}|\Delta \M{P})f(N_{\mathrm{NA}}|\Delta \M{P}). 
\label{eq:proof2}
\vspace{.7mm}
\end{align}
Note that \ref{assumption:indepdentTheta} implies
\begin{equation}
f(\V{\theta}|\Delta \M{P}) =
 \left( \prod_{i\in \Set{D}_{\V{a}}}  f_{\mathrm{A}}(\Delta\M{P})\right) \prod_{i'\notin \Set{D}_{\V{a}}}  (1-f_{\mathrm{A}}(\Delta\M{P}))
 \label{eq:proof3}
\end{equation}
and from \ref{assumption:poissonClutter}, we obtain  \vspace{-.8mm}
\begin{equation}
f(N_{\mathrm{NA}}|\Delta \M{P}) = \frac{\lambda_{\mathrm{NA}}^{N_{\mathrm{NA}}}}{N_{\mathrm{NA}}!}e^{-\lambda_{\mathrm{NA}}}.\vspace{0mm}
 \label{eq:proof4}
\end{equation}
Let us denote by $\Set{A}$ the set of all vectors $\V{a}$ that do not violate assumption A1).  From basic results of combinatorics and \ref{assumption:uniformAssociations}, it is straightforward to show that $f(\V{a}|\V{\theta},N_{\mathrm{NA}},\Delta \M{P}) = N_{\mathrm{NA}}! / N_{\mathrm{S}}!$ for $\V{a} \in \Set{A}$ and thus $f(\V{a}|\V{\theta},N_{\mathrm{NA}},\Delta \M{P}) = \Psi(\V{a})$ $N_{\mathrm{NA}}! / N_{\mathrm{S}}! $ for arbitrary $\V{a} \in \{0,1,\dots,N_{\mathrm{S}}\}^{N_{\mathrm{D}}}$. Using this result as well as  \eqref{eq:proof3} and \eqref{eq:proof4} in \eqref{eq:proof2}, we obtain 
\begin{align}
\hspace{0mm}f(\V{a},N_{\mathrm{S}}|\Delta \M{P}) &=  \Psi(\V{a})  \hspace{.3mm} \frac{\lambda_{\mathrm{NA}}^{N_{\mathrm{S}}- |\Set{D}_{\V{a}}|}}{N_{\mathrm{S}}!}\hspace{-.2mm} e^{-\lambda_{\mathrm{NA}}} \hspace{-.4mm} \left( \hspace{.4mm} \prod_{i\in \Set{D}_{\V{a}}} \hspace{-.5mm} f_{\mathrm{A}}(\Delta\M{P})   \hspace{-.3mm}\right) \nn\\[.5mm]
&\hspace{10mm}\times \prod_{j\notin \Set{D}_{\V{a}}}  \big(1-f_{\mathrm{A}}(\Delta\M{P})  \big).
\label{eq:proof2a}
\vspace{.7mm}
\end{align}
Furthermore, using  \vspace{-1mm} \ref{assumption:indepdentAssoNonAssoMeasurements}-\ref{assumption:clutterDistribution} $f(\V{z}|\V{a},N_{\mathrm{S}},\Delta \M{P})$ in \eqref{eq:conditionalPDFDer}, can be expressed \vspace{0mm}as
\begin{align}
&f(\V{z}|\V{a},N_{\mathrm{S}},\Delta \M{P}) \nn\\
&\hspace{4mm}= \left( \hspace{.4mm} \prod_{j=1}^{N_{\mathrm{S}}} f_{\mathrm{NA}}(\V{s}_{\mathrm{H}}^{(j)}|\Delta\M{P})  \right) \prod_{i\in \mathcal{D}_{\V{a}}} \frac{f\big(\V{z}^{(i,a^{(i)})}|\Delta\M{P}\big)}{f_{\mathrm{NA}}\big(\V{s}_{\mathrm{H}}^{(a^{(i)})}\big|\Delta\M{P}\big)}. \label{eq:likelihood}\\[-4.5mm]
\nn
\end{align}

Now, we use the expression for $f(\V{a},N_{\mathrm{S}}|\Delta \M{P})$ in  \eqref{eq:proof2a} and the expression for $f(\V{z}|\V{a},N_{\mathrm{S}},\Delta \M{P})$ in \eqref{eq:likelihood} into \eqref{eq:proof2} and obtain
\begin{align}
&f(\V{z},\V{a},N_{\mathrm{S}}|\Delta \M{P}) \nn\\[.5mm]
&\hspace{1mm}=  \Psi(\V{a})  \hspace{.3mm} \frac{\lambda_{\mathrm{NA}}^{N_{\mathrm{S}}- |\Set{D}_{\V{a}}|}}{N_{\mathrm{S}}!}\hspace{.3mm} e^{-\lambda_{\mathrm{NA}}} \left( \hspace{.4mm} \prod_{j=1}^{N_{\mathrm{S}}} f_{\mathrm{NA}}(\V{s}_{\mathrm{H}}^{(j)}|\Delta\M{P})  \right) \nn\\[.5mm]
&\hspace{1mm}\times \left( \prod_{i\in \mathcal{D}_{\V{a}}} \frac{f_{\mathrm{A}}(\Delta\M{P})  \hspace{.3mm} f\big(\V{z}^{(i,a^{(i)})}\big|\Delta\M{P}\big)}{f_{\mathrm{NA}}\big(\V{s}_{\mathrm{H}}^{(a^{(i)})}\big|\Delta\M{P}\big)}  \right) \prod_{j\notin \Set{D}_{\V{a}}}  \big(1-f_{\mathrm{A}}(\Delta\M{P})  \big). \nn\\[-1.5mm]
\label{eq:finalAppendix} \\[-3mm]
\nn
\end{align}
Finally, the final expression for $f(\V{z},\V{a},\V{b},N_{\mathrm{S}}|\Delta \M{P})$ in \vspace{0mm} \eqref{eq:conditionalPDF} is obtained from the expression for $f(\V{z},\V{a},N_{\mathrm{S}}|\Delta \M{P})$ in \eqref{eq:finalAppendix} by ``opening'' the factor $\Psi(\V{a})$. In particular, we introduce $\V{b}$ as described in \cite[Section V.B]{MeyerKropfreiterWilliamsLauHlawatschBracaWin:J18} and replace $\Psi(\V{a})$ by $\Psi(\V{a},\V{b})$ in \eqref{eq:psiAB}.
\renewcommand{\baselinestretch}{1}
\selectfont
\bibliographystyle{IEEEtran}
\bibliography{IEEEabrv,StringDefinitions,mendrzikCvLib,allReferencesLib}

\begin{thebibliography}{10}
\providecommand{\url}[1]{#1}
\csname url@samestyle\endcsname
\providecommand{\newblock}{\relax}
\providecommand{\bibinfo}[2]{#2}
\providecommand{\BIBentrySTDinterwordspacing}{\spaceskip=0pt\relax}
\providecommand{\BIBentryALTinterwordstretchfactor}{4}
\providecommand{\BIBentryALTinterwordspacing}{\spaceskip=\fontdimen2\font plus
\BIBentryALTinterwordstretchfactor\fontdimen3\font minus
  \fontdimen4\font\relax}
\providecommand{\BIBforeignlanguage}[2]{{%
\expandafter\ifx\csname l@#1\endcsname\relax
\typeout{** WARNING: IEEEtran.bst: No hyphenation pattern has been}%
\typeout{** loaded for the language `#1'. Using the pattern for}%
\typeout{** the default language instead.}%
\else
\language=\csname l@#1\endcsname
\fi
#2}}
\providecommand{\BIBdecl}{\relax}
\BIBdecl

\bibitem{Ziebinski:C16}
A.~Ziebinski, R.~Cupek, H.~Erdogan, and S.~Waechter, ``A survey of {ADAS}
  technologies for the future perspective of sensor fusion,'' in
  \emph{Computational Collective Intelligence}.\hskip 1em plus 0.5em minus
  0.4em\relax Springer International Publishing, 2016, pp. 135--146.

\bibitem{MalliosRidaoRibasMaurelliPetillot:C10}
A.~{Mallios}, P.~{Ridao}, D.~{Ribas}, F.~{Maurelli}, and Y.~{Petillot},
  ``{EKF-SLAM for AUV navigation under probabilistic sonar scan-matching},'' in
  \emph{Proc. IEEE IROS-10}, Taipei, Taiwan, 2010, pp. 4404--4411.

\bibitem{BiberStrasser:C03}
P.~Biber and W.~Strasser, ``The normal distributions transform: a new approach
  to laser scan matching,'' in \emph{Proc. IEEE IROS-03}, Las Vegas, NV, USA,
  Oct. 2003.

\bibitem{ZhangSingh:J16}
J.~Zhang and S.~Singh, ``Low-drift and real-time lidar odometry and mapping,''
  \emph{Auton. Robots}, vol.~41, no.~2, pp. 401--416, Feb. 2016.

\bibitem{SrivatsanXuZevallosChoset:J18}
R.~A. Srivatsan, M.~Xu, N.~Zevallos, and H.~Choset, ``{Probabilistic pose
  estimation using a Bingham distribution-based linear filter},'' \emph{Int J.
  Robot Res.}, vol.~37, no. 13-14, pp. 1610--1631, Jun. 2018.

\bibitem{Deschaud:C18}
J.-E. Deschaud, ``{IMLS}-{SLAM}: Scan-to-model matching based on {3D} data,''
  in \emph{Proc. IEEE ICRA-18}, Brisbane, Australia, May 2018.

\bibitem{ShanEnglot:C18}
T.~Shan and B.~Englot, ``{LeGO}-{LOAM}: Lightweight and ground-optimized lidar
  odometry and mapping on variable terrain,'' in \emph{Proc. IEEE IROS-18},
  Madrid, Spain, Oct. 2018.

\bibitem{TangChenNiuWangChenLiuShiHyyppae:J15}
J.~Tang, Y.~Chen, X.~Niu, L.~Wang, L.~Chen, J.~Liu, C.~Shi, and J.~Hyyppä,
  ``{LiDAR} scan matching aided inertial navigation system in {GNSS}-denied
  environments,'' \emph{Sensors}, vol.~15, no.~7, pp. 16\,710--16\,728, Jul.
  2015.

\bibitem{Olson:C09}
E.~Olson, ``Real-time correlative scan matching,'' in \emph{2009 {IEEE}
  International Conference on Robotics and Automation}.\hskip 1em plus 0.5em
  minus 0.4em\relax {IEEE}, may 2009.

\bibitem{MalliosRidaoRibasHernandez:J13}
A.~Mallios, P.~Ridao, D.~Ribas, and E.~Hern{\'{a}}ndez, ``Scan matching {SLAM}
  in underwater environments,'' \emph{Autonomous Robots}, vol.~36, no.~3, pp.
  181--198, jun 2013.

\bibitem{BarShalomWillettTian:B11}
Y.~Bar-Shalom, P.~K. Willett, and X.~Tian, \emph{{Tracking and Data Fusion: A
  Handbook of Algorithms}}.\hskip 1em plus 0.5em minus 0.4em\relax Storrs, CT,
  USA: Yaakov Bar-Shalom, 2011.

\bibitem{WilliamsLau:J14}
J.~L. Williams and R.~Lau, ``Approximate evaluation of marginal association
  probabilities with belief propagation,'' \emph{{IEEE} Trans. Aerosp.
  Electron. Syst.}, vol.~50, no.~4, pp. 2942--2959, Oct. 2014.

\bibitem{MeyerWin:J20}
F.~{Meyer} and M.~Z. {Win}, ``Scalable data association for extended object
  tracking,'' \emph{IEEE Trans. Signal Inf. Process. Netw.}, vol.~6, pp.
  491--507, 2020.

\bibitem{MeyWil:C20}
F.~{Meyer} and J.~L. {Williams}, ``Scalable detection and tracking of extended
  objects,'' in \emph{Proc. IEEE ICASSP 2020}, May 2020, pp. 8916--8920.

\bibitem{ThrunBurgardFox:B05}
S.~Thrun, W.~Burgard, and D.~Fox, \emph{Probabilistic Robotics}.\hskip 1em plus
  0.5em minus 0.4em\relax MIT Press Ltd, 2005.

\bibitem{CarrBeatsonCherrieMitchellFrightMcCallumEvans:C01}
J.~C. Carr, R.~K. Beatson, J.~B. Cherrie, T.~J. Mitchell, W.~R. Fright, B.~C.
  McCallum, and T.~R. Evans, ``Reconstruction and representation of {3D}
  objects with radial basis functions,'' in \emph{Proc. ACM {SIGGRAPH}-01}, Los
  Angeles, CA, USA, Aug. 2001.

\bibitem{DemantkeMalletDavidVallet:J12}
J.~Demantk{\'{e}}, C.~Mallet, N.~David, and B.~Vallet, ``Dimensionality based
  scale selection in {3D} lidar point clouds,'' \emph{ISPRS Archives}, vol.
  {XXXVIII}-5/W12, pp. 97--102, Sep. 2012.

\bibitem{KlasingAlthoffWollherrBuss:C09}
K.~Klasing, D.~Althoff, D.~Wollherr, and M.~Buss, ``Comparison of surface
  normal estimation methods for range sensing applications,'' in \emph{Proc.
  IEEE ICRA-09}, Kobe, Japan, May 2009.

\bibitem{Low:TR04}
K.-L. Low, ``Linear least-squares optimization for point-to-plane {ICP} surface
  registration,'' Department of Computer Science, University of North Carolina
  at Chapel Hill, Tech. Rep., 2004.

\bibitem{Richards:C95}
P.~Richards, ``{Constrained Kalman filtering using pseudo-measurements},'' in
  \emph{{IEE} CATT-95}, London, UK, May 1995.

\bibitem{LoeligerDauwelsHuKorlPingKschischang:J07}
H.-A. Loeliger, J.~Dauwels, J.~Hu, S.~Korl, L.~Ping, and F.~R. Kschischang,
  ``The factor graph approach to model-based signal processing,'' \emph{Proc.
  {IEEE}}, vol.~95, no.~6, pp. 1295--1322, Jun. 2007.

\bibitem{Loeliger:J04}
H.~Loeliger, ``An introduction to factor graphs,'' \emph{{IEEE} Signal Process.
  Mag.}, vol.~21, no.~1, pp. 28--41, Jan. 2004.

\bibitem{KschischangFreyLoeliger:J01}
F.~Kschischang, B.~Frey, and H.-A. Loeliger, ``Factor graphs and the
  sum-product algorithm,'' \emph{IEEE Trans. Inf. Theory}, vol.~47, no.~2, pp.
  498--519, 2001.

\bibitem{MeyerKropfreiterWilliamsLauHlawatschBracaWin:J18}
F.~Meyer, T.~Kropfreiter, J.~L. Williams, R.~Lau, F.~Hlawatsch, P.~Braca, and
  M.~Z. Win, ``Message passing algorithms for scalable multitarget tracking,''
  \emph{Proc. {IEEE}}, vol. 106, no.~2, pp. 221--259, Feb. 2018.

\bibitem{DoucetFreitasGordon:B01}
A.~Doucet, N.~de~Freitas, and N.~Gordon, \emph{{Sequential Monte Carlo Methods
  in Practice}}.\hskip 1em plus 0.5em minus 0.4em\relax New York, NY, USA:
  Springer, 2001.

\bibitem{Boumal:B20}
\BIBentryALTinterwordspacing
N.~Boumal, ``An introduction to optimization on smooth manifolds,'' Published
  online, Aug 2020. [Online]. Available:
  \url{http://www.nicolasboumal.net/book}
\BIBentrySTDinterwordspacing

\bibitem{TchapmiChoyArmeniGwakSavarese:C17}
L.~Tchapmi, C.~Choy, I.~Armeni, J.~Gwak, and S.~Savarese, ``{SEGCloud}:
  Semantic segmentation of 3d point clouds,'' in \emph{2017 International
  Conference on 3D Vision (3DV)}.\hskip 1em plus 0.5em minus 0.4em\relax
  {IEEE}, oct 2017.

\end{thebibliography}

\end{document}